\documentclass[
aps,%
12pt,%
final,%
notitlepage,%
oneside,%
onecolumn,%
nobibnotes,%
nofootinbib,%
superscriptaddress,%
noshowpacs,%
centertags]%
{revtex4}

\def\la{\mathrel{\mathchoice {\vcenter{\offinterlineskip\halign{\hfil
$\displaystyle##$\hfil\cr<\cr\sim\cr}}}
{\vcenter{\offinterlineskip\halign{\hfil$\textstyle##$\hfil\cr
<\cr\sim\cr}}}
{\vcenter{\offinterlineskip\halign{\hfil$\scriptstyle##$\hfil\cr
<\cr\sim\cr}}}
{\vcenter{\offinterlineskip\halign{\hfil$\scriptscriptstyle##$\hfil\cr
<\cr\sim\cr}}}}}

\def\utw{\smash{\rlap{\lower5pt\hbox{$\sim$}}}}
\def\udtw{\smash{\rlap{\lower6pt\hbox{$\approx$}}}}

\def\diameter{{\ifmmode\mathchoice
{\ooalign{\hfil\hbox{$\displaystyle/$}\hfil\crcr
{\hbox{$\displaystyle\mathchar"20D$}}}}
{\ooalign{\hfil\hbox{$\textstyle/$}\hfil\crcr
{\hbox{$\textstyle\mathchar"20D$}}}}
{\ooalign{\hfil\hbox{$\scriptstyle/$}\hfil\crcr
{\hbox{$\scriptstyle\mathchar"20D$}}}}
{\ooalign{\hfil\hbox{$\scriptscriptstyle/$}\hfil\crcr
{\hbox{$\scriptscriptstyle\mathchar"20D$}}}}
\else{\ooalign{\hfil/\hfil\crcr\mathhexbox20D}}%
\fi}}

\begin{document}

\title{A Search for Small-Scale Clumpiness
in Dense Cores of Molecular Clouds}

\author{\firstname{L.~E.}~\surname{Pirogov}} 
\email{pirogov@appl.sci-nnov.ru}
\affiliation{%
Institute of Applied Physics Russian Academy of Sciences
}%
\author{\firstname{I.~I.}~\surname{Zinchenko}}
\affiliation{%
Institute of Applied Physics Russian Academy of Sciences
}%

\begin{abstract}
We have analyzed HCN(1--0) and CS(2--1) line profiles obtained 
with high signal-to-noise ratios toward distinct positions 
in three selected objects in order to search for small-scale structure in
molecular cloud cores associated with regions of high-mass star formation. 
In some cases, ripples were detected in the line profiles, which could be due 
to the presence of a large number of unresolved small clumps
in the telescope beam. 
The number of clumps for regions with linear scales of $\sim 0.2-0.5$~pc 
is determined using an analytical model and detailed calculations 
for a clumpy cloud model; this number varies in the
range: $\sim 2\times 10^4-3\times 10^5$, depending on the source. 
The clump densities range from $\sim 3\times 10^5-10^6$~cm$^{-3}$,
and the sizes and volume filling factors of the clumps are $\sim(1-3)\times 10^{-3}$~pc 
and $\sim 0.03-0.12$. 
The clumps are surrounded by inter-clump gas with densities 
not lower than $\sim (2-7)\times 10^4$~cm${-3}$. 
The internal thermal energy of the gas in the model clumps is much higher 
than their gravitational energy. 
Their mean lifetimes
can depend on the inter-clump collisional rates, and vary in the range 
$\sim 10^4-10^5$~yr. 
These structures are probably connected with density fluctuations 
due to turbulence in high-mass star-forming regions.

{Key words:} interstellar medium, molecular clouds,
interstellar molecules, radio lines.

\end{abstract}

\maketitle

\section{INTRODUCTION}

The problem of why star formation results in the
formation of stars in clusters in some cases and in
the appearance of isolated stars in others remains
unsolved. It is known that stellar clusters in which
high-mass stars are born are associated with higher
mass, more turbulent cores of molecular clouds than
those in which isolated stars are born. Regions of
molecular emission associated with such cores probably
do not fill the telescope beam completely; instead,
they may have an inhomogeneous, clumpy structure
that is unresolved in the observations (see, e.g., [1]).
There are many indirect pieces of evidence suggesting
the existence of small-scale (unresolved) inhomogeneities
in the dense cores of interstellar molecular
clouds that are birthplaces of high-mass stars and
stellar clusters. In particular, a clumpy structure for
the molecular emission in objects associated with
H~II regions and H$_2$O masers follows from the ratios
of the observed peak intensities and widths of lines
with different optical depths (see, e.g., [2, 3]); this
is confirmed by modeling of line profiles for various
rotational transitions of the CS molecule in models
with small clumps [4]. The absence of appreciable
density variations in giant molecular clouds, where
the gas column densities vary by more than an order
of magnitude [5], could also be due to the existence
of small-scale density inhomogeneities. The masses
of the cores derived from calculations of molecular
excitation often prove to be higher than the virial
masses or masses calculated from observational data
on optically thin dust emission, likewise suggesting a
clumpy structure for the molecular emission regions
(see, e.g., [6]). The detection of extended regions of
emission in atomic carbon lines toward a number of
high-mass star-forming regions [7] may also testify to
the inhomogeneous structure of these regions, where
more rarefied photon-dominated regions, where the
effects of ultraviolet radiation are important, can apparently
coexist with high-density gas.

Important evidence for small-scale clumpiness in
high-mass star-forming regions is provided by observations
of anomalies in the hyperfine structure of
the $J=1-0$ HCN line, manifest as a low relative
intensity of the $F=1-1$ component compared to
the optically thin case. This is due to the overlap of
local profiles of components in higher lying transitions,
and depends on the degree of broadening of
the local profiles. These anomalies arise in gas with
kinetic temperatures $\ge 20$~K, and should vanish when
the widths of the local profiles become close to that
of the observed HCN(1--0) profiles in these objects
($\ge 2$~km/s). However, if the objects consist of small,
randomly moving clumps with densities close to the
critical density for the given molecular transition and
the line profiles of individual clumps are close to
thermal, the observed HCN(1--0) lines will display
these anomalies, but have widths corresponding to
the velocity dispersion of their relative motions [8].

Estimation of the physical properties of small
clumps that are not directly resolved in observations
requires detailed calculations of radiative transfer in
an inhomogeneous, fragmentary medium, with the
subsequent computation of line profiles and comparison
with the results of observations. This can involve
a large number of unknown parameters and be fairly
resource-intensive in a full study of the entire space
of possible values of these parameters. Martin et al.
[9] proposed a simple analytical model in which a
cloud was taken to consist of unresolved, randomly
moving identical clumps with a small volume filling
factor. Using this model, we can use the ratios of the
intensities and widths of two lines having different
optical depths to derive the relationships between
parameters of the clumps. It was pointed out in [10,
11] that, if the volume filling factor of the clumps is
low and the number of clumps in the line of sight at
a given radial velocity is small, we should expect the
appearance of ripples in line profiles. These ripples
are due to fluctuations in the number of clumps in
the line of sight at various velocities, and can be used
to estimate the parameters of the clumps filling the
beam.

However, detecting such ripples requires high
sensitivity and high spectral resolution. To check the
method, the above studies used mainly observational
data on the most intense lower transitions of the CO
molecule and its isotopic modifications in the dense
cores of M17~SW and Orion~A. Since these lines are
efficient probes of gas with densities of $\sim 10^3$~cm$^{-3}$,
they are most likely tracers of the inter-clump gas,
as is inferred by the spatial correlation of the atomic
carbon and CO emission (see, e.g., [12]). Searches
for small-scale inhomogeneities are probably more
fruitfully carried out using lines that trace gas with
densities one to two orders of magnitude higher than
the low CO transitions, such as CS(2--1) or HCN(1--0).

In this paper, we present measurements of CS(2--1) and HCN(1--0) lines, 
as well as isotopic modifications
of these lines, toward selected positions in three
dense molecular-cloud cores associated with highmass
star-forming regions and possibly possessing
an inhomogeneous, clumpy structure. We used the
analytical method of [9, 11] and compared the observational
data with the results of detailed calculations
for a simple model of a cloud with small clumps.
Our analysis has enabled us to estimate the physical
properties of small, spatially unresolved clumps in the
framework of the models considered.

\section{RESULTS OF THE OBSERVATIONS}

To confirm the existence of small-scale clumpiness
of the dense gas in high-mass star-forming
regions, we observed three dense cores associated
with high-mass star-forming regions on the 20-m
radio telescope in Onsala (Sweden) in 1999. We
chose for these observations the objects S140, S199
(IC1848A, AFGL 4029), and S235 (G173.72+2.70),
which display anomalies in the hyperfine structure of
the $J=1-0$ HCN line, suggesting the existence of
small clumps with thermal linewidths [8].

The observations were carried out in the $J=1-0$
line of HCN and $J=2-1$ line of CS at 88.63 and
97.98 GHz, respectively. Some positions were observed
in lines of rarer isotopes of these molecules,
namely, the $J=1-0$ line of H$^{13}$CN and $J=2-1$ line
of C34S (at 86.34 and 96.41 GHz, respectively). The
telescope beamwidths were 45$''$ at the HCN(1--0)
frequency and 39$''$ at the CS frequency. The beam 
efficiencies at these frequencies were 0.59 and 0.56, respectively.
The dependence of this factor on the source
elevation was not taken into account; according to
later measurements, this could result in an underestimation
of the line intensities at low elevations of up
to $\sim 30$\%. The single-sideband noise temperature of
the system during most of the observing time was
$\sim 170-350$~K, depending on the weather conditions
and source elevation. The observations were carried
out toward two positions separated by an angular
distance of 1.5$'$--2$'$, which exceeded the telescope
beamwidth; thus, the distributions of small clumps for
these positions could be assumed statistically independent.
The signal-to-noise ratios were $\sim 30-250$,
and the integration times were 1--11~h, depending on
the source and observed line. The list of sources, their
coordinates and distances, and the spatial resolution
of the observations in the various lines are given in
Table 1.

The spectrum analyzer was an autocorrelator with
a 20-MHz bandwidth and a frequency resolution of
12.5 kHz; this corresponded to a velocity resolution of
$\sim 40$~m/s at the observed frequencies. However, analysis
of the spectra at emission-free reference positions
showed that adjacent channels of the autocorrelator
were not independent (as was also noted in [11]),
which degraded the actual spectral resolution. The
correlation coefficient between adjacent channels was
$\sim 0.6$, between channels $i$ and $i+2$ was $\sim 0.2-0.3$,
and between channels $i$ and $i+3$ was close to zero.
Accordingly, only one of three successive channels
was used in the subsequent analysis; thus, the spectral
resolution in velocity was 127 m/s.

Figure 1 shows the spectra obtained. The observed
profiles demonstrate no self-absorption effects. In a
number of cases, the profiles of themain-isotope lines
(HCN, CS) obtained with high signal-to-noise ratio
display a small asymmetry, which is noticeable during
a comparison with the centers of the lines of rarer
isotopes. High-velocity wings due to the presence of
large-scalemotions in the sources are also present. In
S235, the HCN and CS profiles have more than one
component with similar intensities and linewidths.

\section{ANALYSIS OF PARAMETERS
OF THE OBSERVED LINE PROFILES
USING AN ANALYTICAL MODEL}

Martin et al. [9] derived an analytical expression
for the profile of a line emitted by a cloud consisting
of identical clumps moving randomly in space.
Assuming that the excitation temperature of the line
(Tex) is identical in all the clumps and neglecting
the contribution from the microwave background, the
radiation temperature of the line is described by the
expression:

\begin{equation}
T_{\rm R}(v)=T_{\rm EX}(1-e^{-\tau_{\rm eff}(v)})\hspace{2mm}.
\label{eq:trad}
\end{equation}

The effective optical depth of a cloud ($\tau_{\rm eff}$) when the
velocity dispersion of the clumps ($\sigma$) is much greater
than that of the gas inside the clumps ($v_0$) can be
written

\begin{equation}
\tau_{\rm eff}(v)=N_c A(\tau_0) \exp(-\pi v^2/\sigma^2)\hspace{2mm},
\end{equation}

\noindent{where $A(\tau_0)$ is an integral function that depends on
the optical-depth distribution and geometry of the
clump, and $\tau_0$ is the maximum optical depth of the clump.}
When $\tau_0\ll 1$, we have $A(\tau_0)\simeq \tau_0$. 
The quantity $N_c$ is the number of clumps in a column with
cross-sectional area $r_0^2$ ($r_0$ is the clump radius) whose
velocities are within v0 of the line center:

\begin{equation}
N_c=K N_{\rm tot} \frac{r_0^2}{B} \frac{v_0}{\sigma}\hspace{2mm},
\end{equation}

\noindent{where $B$ is the area of the telescope beam and $N_{\rm tot}$
is the number of clumps in the beam.} The factor
$K$ depends on the optical-depth distribution in an
individual clump. Martin et al. [9] and Tauber [11]
assumed a Gaussian dependence for the optical depth
on the impact parameter, in which case $K=1$. In the
case of opaque disks, when the optical depth does not
depend on the impact parameter, $K=\pi$.

Martin et al. [9] used ratios of the peak intensities
and widths of two lines (lower transitions of CO and
$^13$CO) to find relationships between physical parameters
of the clumps in the dense core M17~SW. Tauber
[11] proposed an analytical method for determining
the clump parameters based on this idea. In addition
to the ratios of the peak intensities and widths of
two lines, a new parameter was introduced (assuming that $N_c\la 10$):

\begin{equation}
\frac{\Delta T_{\rm R}}{T_{\rm R}}=
\frac{\tau_{\rm eff}}
{(e^{\tau_{\rm eff}}-1) \sqrt{N_c\frac{B}{r{_0}^2}}} =
\frac{\tau_{\rm eff}}
{(e^{\tau_{\rm eff}}-1)\sqrt{N_{\rm tot} \frac{v_0}{\sigma}}} \hspace{2mm},
\label{eq:smooth}
\end{equation}

\noindent{where $\Delta T_{\rm R}$ is the standard deviation for fluctuations
of the radiation temperature in some range near the
line center relative to a certain expected value; these
deviations are related to fluctuations in the number
of clumps in the line of sight moving with various
velocities.} The more rugged and less smooth the line
profile, the greater the ratio $\Delta T_{\rm R}/T_{\rm R}$; henceforth, we
will call this ratio the ``jaggedness" of the line profile,
in contrast to the term ``smoothness" used in [11].

The values of $T_{\rm R}$, $\Delta T_{\rm R}$ and the widths for
two lines with different optical depths can be determined
from observations. If one of the lines is optically
thin, we can calculate the value of $\tau_{\rm eff}$ for an
optically thick line from the linewidth ratio. Knowing
the jaggedness and effective optical depth of the line,
we can calculate $N_{\rm tot} \frac{v_0}{\sigma}$ using (4). 
Assuming that the
velocity dispersion in the clumps is determined by the
thermal gas motions and knowing the kinetic temperature,
we can estimate the ratio $\frac{v_0}{\sigma}$
and the number of clumps in the telescope beam. If the jaggednesses
and linewidth ratios for two lines with different optical
depths are known, we can calculate $N_c$ and $\tau_0$ for one
of the lines and estimate the size of a single clump $r_0$.

Martin et al. [9] and Tauber [11] also used an
expression for the ratio of the peak intensities of two
lines: 
$T^1_{\rm R}/T^2_{\rm R}=(1-\exp(-\tau^1_{\rm eff}))/(1-\exp(-\tau^2_{\rm eff}))$,
which is applicable for lines in LTE (e.g., for lower
transitions of CO and $^{13}$CO). However, it is not
appropriate to use this relation for lines of HCN,
CS, and their isotopes, for which the distribution
of population densities over the rotational levels is
appreciably nonequilibrium for the densities and column
densities typical of the dense cores of interstellar
clouds.

As Tauber [11] pointed out, $\Delta T_{\rm R}$ can be estimated
by fitting and subtracting a Gaussian function,
a function of the type (1), or triplets with a
fixed separation between their components (in the
case of HCN); fitting polynomial functions to separate
segments of the line profile, with subsequent
subtraction; or filtration of the lowest harmonics of
the Fourier spectrum, which correspond to the main
line profile. After applying these methods, the residual
spectrum should be noise-like, possibly with different
dispersions within and outside the line. Since the
contribution of the emission of a set of small clumps
is statistically independent from the atmospheric and
instrumental noise in the spectra, we can calculate
the standard deviation for the residual noise: 
$\Delta T_{\rm R}=\sqrt{\Delta T_{\rm L}^2-\Delta T_{\rm N}^2}$,
where $\Delta T_{\rm L}$ is the standard deviation
for temperature fluctuations near the line peak, and
$\Delta T_{\rm N}$ is the standard deviation for fluctuations outside
the line. In his analysis of line parameters for the
spectrum of Orion~A, Tauber [11] fit and subtracted
a function of the form (1); however, the similarity of
the asymmetry in the line profiles obtained toward
different positions of the Orion A cloud led him to
suspect that there was a spatial correlation of the
residual fluctuations, which could be due to systematic
motions. This contradicted the idea that the fluctuations
related to the small-scale clumpiness were
statistically independent, and prevented Tauber [11]
from estimating the jaggedness of the profiles. No
other methods for estimating $\Delta T_{\rm R}$ were applied in [11].

In the calculations of $\Delta T_{\rm L}$, the three methods
mentioned above were applied to all the derived pro-
files. If no excess of $\Delta T_{\rm L}$ above $\Delta T_{\rm N}$ 
was found in at least one of them, it was further assumed that there
were no residual fluctuations related to clumpiness
in this profile. When calculating $\Delta T_{\rm L}$ by fitting and
subtracting polynomials, the bandwidth near the line
peak (2--3 km/s) and the order of the fitted polynomial
(second to fourth) were varied. The optimal cutoff
boundary was chosen for rejecting the lowest Fourier
harmonics. Though the main order of the Fourier
harmonics corresponding to the observed lines with
widths of $\ge 2$~km/s is concentrated at reciprocal velocities
$\le 0.5$~(km/s)$^{-1}$, the presence of even a small
asymmetry in the line profiles of the main isotopes
(possibly due to systematic motions in the dense gas)
can result in the appearance of features in the power
spectra in a broader band than the range corresponding
to the effective Gaussian profile, as can be seen
from Fig.~2 in the case of S140(0,0). We modeled
the emission of a multilayer medium in order to study
the effect of systematic line-of-sight motions on the
profile of a single line or several nearby lines with
optical depths $\ge 1$; each of layers had its own lineof-
sight velocity and excitation temperature, in accordance
with specified power laws. We were able to
achieve profile asymmetries similar to those observed
by varying the velocity at the boundary and the exponent
in the radial dependence of the velocity, as well
as the optical depth of an individual layer. Analysis of
the power spectra for lines with widths $\ge 2$~km/s and
with the addition of noise similar to the measurement
noise showed that, in Gaussian-like spectra, features
associated with systematic motions fall to the noise
level on scales $V^{-1}\le 0.7$~(ª¬/á)$^{-1}$. Therefore, the
cutoff boundary for rejecting Fourier harmonics was
set to 0.7~(ª¬/á)$^{-1}$ (for the HCN(1--0) profiles in
S235, 0.9~(ª¬/á)$^{-1}$). This slightly underestimates
$\Delta T_{\rm L}$, whereas fitting smooth functions to the line
profiles or profile segments may yield overestimates.
We adopted the standard deviation of the temperature
fluctuations in the channels after subtracting the
baseline outside the spectral line as $\Delta T_{\rm N}$.

The peak intensities, linewidths, and effective
optical depths are listed in Table~2. The values of
$T_{\rm R}$ and $\Delta V$ were obtained by fitting Gaussians to
the observed line profiles. For the HCN(1--0) and
H$^{13}$CN(1--0) profiles, these values refer to the central
component $F=2-1$; the relative intensities $R_{\rm 12}$ and
$R_{\rm 02}$ are given for the side components, $F=1-1$
and $F=0-1$. In several cases, we list the widths
of the $F=0-1$ component, which were used when
calculating $\tau_{\rm eff}$. In the remaining cases, $\tau_{\rm eff}$ was
calculated by comparing the linewidths for the main
and rarer isotopes. The errors in the line intensities
and widths were obtained from the fits.

Figure~2 presents the HCN(1--0) and CS(2--1)
line profiles toward some observed positions together
with the residual noise obtained after rejecting the
lowest Fourier harmonics. For each profile, the power
spectrum for small amplitudes of the Fourier harmonics
is shown at the right; the cutoff boundary is also
indicated.

Fluctuations in the residual noise near a line
peak were detected most confidently for the $F=2-1$
HCN(1--0) component. In two cases, fluctuations
of the residual noise were detected in the CS(2--1)
profiles. No fluctuations exceeding the noise level
outside the line were found in the C$^{34}$S(2--1) and
H$^{13}$CN(1--0) profiles; thus, were we able to estimate
only the total number of clumps in the telescope
beam $N_{\rm tot}$, using themethod described above. 
Table~3 lists the values of $\Delta T_{\rm N}$
and $\Delta T_{\rm R}$. The errors of $\Delta T_{\rm R}$
probably reach $\sim 10-20$\%; they are mainly due to our
choice of the interval near the line peak for which
they are calculated, as well as to the computational
method. The table also lists the kinetic temperatures
required to calculate the gas velocity dispersion inside
the clumps, together with the values of $N_{\rm tot}$. Since
uncertainties in $\Delta T_{\rm R}$ and $\tau_{\rm eff}$ strongly influence 
$N_{\rm tot}$, probably resulting in errors no less than $\sim 50$\%, we
retained only two significant figures in these results.

The estimates of $N_{\rm tot}$ based on the HCN(1--0) and
CS(2--1) data toward S140(0,0) differ appreciably;
this could be due to inter-clump gas resulting in
effective smoothing of the fluctuations in the CS(2--1) profiles. 
This possibility is confirmed by model
calculations (see Section~4). In the remaining cases,
the estimates based on $\Delta T_{\rm R}$ for these two lines are
consistent with each other within the probable errors.
For S199(0,0), the value of $\tau_{\rm eff}$ for both lines was
calculated from the ratio of the widths of the 
$F=2-1$ and $F=0-1$ HCN(1--0) components. However,
the estimate of $N_{\rm tot}$ based on this ratio proved to be
considerably lower than the value yielded by themodel
calculations (see Section 4). This may suggest that
this method has overestimated $\tau_{rm eff}$. Toward both positions
in S235, the line profiles contain features that
could be due to the presence of an individual dense
clump in the line of sight is larger than the telescope
beam. Its effect is appreciabe in the HCN(1--0) profile
toward position (0,0), and it is especially obvious toward
position (0,--2) in the CS(2--1) and HCN(1--0)
profiles (Fig.~1). Fitting two Gaussians to the CS(2--1) profile 
and two triplets to the HCN(1--0) profile
toward position (0,--2) enabled us to distinguish two
components with similar widths at velocities --17 and
--16 km/s. The profiles of the $F=2-1$ HCN(1--0)
hyperfine component and the CS(2--1) line toward
position (0,0) are smoother, with only a slight asymmetry.
The estimates $N_{\rm tot}$ were made for the position
(0,0); $\tau_{\rm eff}$ for both lines was calculated from the ratio
of the CS and C$^{34}$S linewidths.

\section{RESULTS OF NUMERICAL
CALCULATIONS IN THE CLUMPY MODEL}

Since no residual fluctuations exceeding the noise
level were found in the profiles of the optically thin
C$^{34}$S(2--1) and H$^{13}$CN(1--0) lines, we were not able
to apply the analytical model (Section 3) to determine
the physical properties of individual clumps. Instead,
we carried out detailed calculations of the excitation
of molecules in a cloud model with small clumps.
We applied the model described in [8], which was
used earlier to explain features in HCN(1--0) spectra
observed in clouds associated with high-mass starforming
regions, in the calculations. An analysis of
the line profile residual intensity fluctuations had not
been conducted earlier. For convenience when comparing
with the analytical model and to reduce the
number of model parameters, we used a simplified
version of the model (Model~1). The cloud was assumed
to be spherically symmetric, isothermal, and
to consist of small cells. Each cell could either be
filled with gas (a clump) or be empty, in accordance
to a specified probability, which has the meaning of
a volume filling factor. The gas density in all clumps
was taken to be identical. It was assumed that the
clumps have no internal structure and move relative
to each other with random velocities having a Gaussian
distribution. The line profile from each individual
clump was assumed to be Gaussian and to have
thermal width. The molecular excitation was calculated
for only one representative clump at the cloud
center, with all the remaining clumps considered to
be identical to it; this matches the conditions of the
analytical model.

We must adjust the temperature, density, abundance
of molecules in a clump, and dispersion of
the relative motion of the clumps to obtain model
profiles that are similar to the observed profiles. The
kinetic temperature of the analyzed sources can be
considered to be known (Table~3), and the dispersion
of the relative motion of the clumps for lines with
moderate optical depth ($\sim 1$) can be estimated from
their widths. The density and abundance of molecules
can be determined fromthe intensities of the 
HCN(1--0) hyperfine structure, which also depend on the ratio
of the filling factor to the size of a single clump. As
an example, Figure~3 shows contours for the intensity
of the $F=2-1$ HCN(1--0) component and the
relative intensities of the side $F=1-1$ and $F=0-1$
components for a kinetic temperature of 30~K and a
velocity dispersion of 1.3~km/s as functions of density
and HCN column density. The ratio of the volume
filling factor to the relative size of a clump (i.e., the
ratio of the clump size to the cloud size) is eight and
four, respectively, for the left-hand and right-hand
graphs; this can reproduce the observed HCN(1--0)
profiles in S140 (0,0) and S199 (0,0), respectively.We
can roughly estimate density and HCN column density
using the observed intensities for the HCN(1--0)
components. Knowing the density and temperature
and varying the abundances of the H$^{13}$CN, CS, and
C$^{34}$S molecules, we can readily obtain line profiles for
these molecules similar to the observed profiles.

Our calculations yielded parameters of the line
profiles that are similar to the observed ones (except
for the high-velocity wings and asymmetry), including
the values of the residual fluctuations
$\Delta T_{\rm R}$ (Table~3). We added synthetic noise with a dispersion
equal to the dispersion of the observed noise outside
the line to the model profiles. The values of $\Delta T_{\rm R}$ were
calculated after rejecting the lowest Fourier harmonics
($\le 0.7$~(km/s)$^{-1}$) for a 3-km/s interval symmetric
about the line center. Since the spatial and velocity
distributions of the clumps (which affect $\Delta T_{\rm R}$) depends
on the initial value of the random-number generator,
we varied the initial values in the calculations
and the results were then averaged. The resulting
errors in the size of a single clump, volume filling
factor, and $N_{\rm tot}$ can reach 20--40\%. The spectral
resolution was set equal to that of the observations.
The probabilities for collisional transitions between
the CS rotational levels were taken from [18]. The
probabilities for collisional transitions between HCN
hyperfine levels were calculated using the method
[19]. A $k$-fold decrease in the size of a clump and
in the volume filling factor together with a $k$-fold
increase in the abundance of molecules in a clump
will keep the total column density and component
intensities constant, and reduce the jaggedness of
the profiles; therefore, by varying these parameters
appropriately, we can change the magnitudes of the
residual fluctuations near the line peak, fitting them to
the observed values.We modeled the HCN(1--0) and
CS(2--1) profiles for both positions in S140 as well as
for S199(0,0), where the observed profile shapes are
close to Gaussian. We did not model the line profiles
in S235 because of their multicomponent structure.

The results of the model calculations are presented
in Table 4. The total number of clumps in the telescope
beam, volume filling factor ($f$), and size ($d$) and
density ($n$) of a single clump were obtained by modeling
the HCN(1--0) profiles using Model 1. The HCN
relative abundances are $10^{-8}$, $4\times 10^{-9}$, and $2.3\times 10^{-9}$ 
for S140(0,0), S140(1.5,0), and S199(0,0), respectively.
The values of $N_{\rm tot}$ are consistent with
the estimates obtained from the analytical model for
both positions in S140. However, the calculations for
S199(0,0) yield a higher value. This is probably due
to overestimation of the effective optical depth in the
HCN line toward this position. It is possible that the
difference in the widths of the $F=0-1$ and $F=2-1$
components, which was the basis for the calculations
of $\tau_{\rm eff}$ in this source, is not due to the optical depth.

However, the model calculations of the 
CS(2--1) profiles show that, for the same clump-structure
parameters as for HCN(1--0), $\Delta T_{\rm R}$ exceeds the observed
values. The largest discrepancy is noted for
both positions in S140. As was pointed out in [11],
taking into account the inter-clump gas can smooth
line-profile fluctuations due to clumpy structure if the
inter-clump density is sufficient to contribute significantly
to the observed lines. This gas can affect the
CS(2--1) profile more appreciably than the HCN(1--0) profile, 
due to the difference in the critical densities
for the excitation of these lines. To bring the model
$\Delta T_{\rm R}$ values for the CS(2--1) profiles into agreement
with the observed ones, we also performed calculations
for a model with inter-clump gas of lower
density (Model~2). The calculations were carried out
in two stages. In the first, we calculated the excitation
of molecules in the inter-clump gas assuming
a homogeneous, isothermal cloud without clumps.
In the second stage, we calculated the excitation of
molecules in clumps, but with the cells, which were
empty in Model~1, now considered to be filled with
inter-clump gas. By varying the density ($n_{\rm ic}$) and
abundance of molecules of the inter-clump gas, we
can obtain profiles with a jaggedness corresponding
to the observed profiles. In the calculations, the
CS abundances in clumps and the inter-clump gas
were identical; the temperature of the inter-clump
gas was set equal to either the temperature of the
gas in clumps or to 200 K, which is fairly characteristic
of photon-dominated regions (see, e.g., [20]).
The velocity dispersion in the inter-clump gas was
taken to be equal to the velocity dispersion of the
clumps, which corresponded to the observed profile
width. The minimum values of the inter-clump gas
density that resulted in effective smoothing of the
CS(2--1) profiles and .TR values in agreement with
the observed ones are listed in the sixth column of
Table 4. The CS relative abundances in the clumps
and inter-clump gas were set to be 
$1.5\cdot 10^{-9}$, $4\cdot 10^{-10}$ and $1.4\cdot 10^{-9}$
for S140(0,0), S140(1.5,0),
and S199(0,0), respectively. However, in order for the
inter-clump gas to have no effect on the parameters
of the HCN(1--0) profiles, we must assume that the
HCN abundance is much lower in this gas than in
clumps. This could come about due to different HCN
formation rates in photon-dominated regions and in
clumps. This problem requires further study.

Figure 4 shows model profiles of the HCN(1--0)
and CS(2--1) lines; these correspond to the observed
profile in S140(0,0), except for the high-velocity
component for HCN(1--0). The CS(2--1) profiles are
given for Models 1 and 2.

Knowing the sizes and densities of model clumps,
we can readily estimate their masses, which proved
to be much lower than the clump masses calculated
from the condition of virial equilibrium (Table~4). The
condition of pressure equilibrium for the clumps is
probably not fulfilled. The ratios of the external and
internal pressures for the clumps in the model with
inter-clump gas (Model~2) are less than unity (Table~4). 
In fact, the pressure ratios could be lower than
the quoted values (by up to a factor of two), due to the
difference between the surface turbulent pressure in
the medium in which the denser sphere is embedded
and the unperturbed pressure far from this region [21].

\section{DISCUSSION}

The analysis of the previous section has shown
that the studied sources can contain a large number
of small, nonequilibrium clumps with densities exceeding
the inter-clump density by an order of magnitude.
We can approximately estimate the lifetime of
an isolated clump in a lower-density medium using
the virial theorem. For an isothermal, spherical clump
subject to external pressure (with the magnetic-field
energy of the clump being negligibe), the virial theorem
can be written

\begin{equation}
\frac{1}{2}\frac{d^2 I}{dt^2}=3\,Mc^2-\alpha\frac{GM^2}{R}-
4\pi R^3\,P_{\rm ext}
\hspace{2mm},
\end{equation}

\noindent{where $I\approx M\,R^2$ is the moment of inertia of the clump,
$M$ its mass, $R$ its radius, $c$ the speed of sound, $\alpha$
factor of the order of unity that depends on the density
distribution (for a uniform distribution, $\alpha=0.6$),
$G$ the gravitational constant, and $P_{\rm ext}$ the external
pressure.} This equation is widely used to estimate
the parameters of clouds and cloud cores in virial
equilibrium ($d^2 I/dt^2=0$). The time during which a
clump with mass M and radius R0 will expand to
radius R1 is

\begin{equation}
t=\frac{R_0}{\sqrt{3}\,c}\int\limits_{1}^{x_1}
\frac{x\,dx}
{\sqrt{(x-x_e)^2-(1-x_e)^2-\frac{2\,P_{\rm ext}}{5\,P_{\rm int}}(x^5-1)}}
\hspace{2mm},
\label{eq:vireq}
\end{equation}

\noindent{where $x_1=R_1/R_0$, $x_e=R_e/R_0$, $R_e=\alpha\,G\,M/3c^2$ is
the radius for which the internal and gravitational
energies are equal, and $P_{\rm int}=3Mc^2/4\pi\,R_0^3$ is the gas
pressure inside a sphere with radius $R_0$.} In [22], a
similar approach was applied to the expansion of a
nonequilibrium, isothermal sphere (without external
pressure) in which the internal energy dominated
the gravitational energy. In this case, the sphere
expands infinitely, and the time required to double its
initial radius ($R_1=2R_0$), and, accordingly, decrease
its density by a factor of eight, was adopted for its
lifetime. The gravitational energy of the considered
model clumps is much less than their internal energy
($x_e$ is $7\cdot 10^{-4}$, $2\cdot 10^{-4}$ ¨
$9\cdot 10^{-3}$ for S140(0,0),
S140(1.5,0), and S199(0,0), respectively). If we set
$P_{\rm ext}=0$ in (6), and neglect $x_e$ compared to $x$ and
unity, the solution is
$t=\sqrt{(R_1^2-R_0^2)/3c^2}$.
In this case, the time for the initial radius to double is equal
to the time for a sound wave to propagate from
the cloud's center to its edge, 
$t_c=R_0/c$, which is $\sim 1.5\cdot 10^3$, $\sim 3.9\cdot 10^3$ and
$t_c\sim 4.5\cdot 10^3$ years for
S140(0,0), S140(1.5,0), and S199(0,0), respectively.

When the pressure of the inter-clump gas is taken
into account, the unlimited expansion of the sphere
is replaced by an oscillation mode, with expansion
followed by contraction and vice versa. The amplitude
of these oscillations can be found from the radicand
in (6), which is real and positive only for values of
$x$ between unity and some largest value ($x_{\rm max}$) that
depends on the pressure ratio. The clumps will expand
until they achieve the maximum radius 
$R_{\rm max}=R_0\cdot x_{\rm max}>R_{\rm eq}$, where $R_{\rm eq}$
is the clump radius when
the pressures come to equilibrium. The internal pressure
of the gas at $R_{\rm max}$ becomes lower than the external
pressure, resulting in contraction. For pressure
ratios $\sim 0.1$, the time when the sphere reaches $R_{\rm max}$
only slightly exceeds $t_c$ (only at $P_{\rm ext}/P_{\rm int}=10^{-3}$
will it reach $\sim 10\,t_c$). For the considered model clumps, the
period of the oscillations is $\sim (1.3-2.8)\,t_c$.

Though this analysis is fairly simplistic, Keto et al.
[23] and Broderick et al. [24] considered the oscillation
mode of a pressure-confined, isolated cloud with internal
thermal motions and without magnetic field in
detail. They found that the lifetime of the oscillations
can be fairly long, comparable to the lifetime of the
cloud, and is mainly determined by radiative losses
and nonlinear interactions between the oscillation
modes. In our model with a large number of identical,
spherical clumps moving at random velocities
relative to each other, lifetimes of the clumps may be
determined by the mean time between collisions, if
clumps are destroyed after a collision. This time is
$L_f/V$, where $L_f$ is the mean free path of a clump
in the cloud and $V$ is the most probable velocity of
the clumps. Supposing that $L_f=(n_f\,d^2)^{-1}$, where $n_f$
is the number density of clumps and $V$ is the velocity
dispersion in the inter-clump gas, we find that
the mean lifetime of the clumps is $\sim 1.7\cdot 10^4$ years
for both the considered regions in S140 
and $\sim 8.4\cdot 10^4$ years for the region toward S199(0,0).

Such short-lived, nonequilibrium structures are
probably fluctuations of density enhancements arising
due to turbulence in a region of formation of highmass
stars and star clusters. If so, our model with
identical, spherical clumps will describe the characteristics
of the fluctuations only approximately. A
more detailed study of the origin and evolution of
small-scale density inhomogeneities in a turbulent
medium requires the use of three-dimensional, magnetohydrodynamic
models together with radiativetransfer
calculations in both lines and continuum.
Important information on the gas structure in regions
of high-mass star formation can be obtained from
future high-sensitivity, high-spectral-resolution observations
in lines of molecules with considerably different
critical densities. The parameters of the smallscale
structure in objects of this class should be provided
by direct observations with a spatial resolution
of $\la 0.1''$, using new-generation interferometers
planned to operate in the next decade (see, e.g., [25]).

\section{CONCLUSIONS}

We carried out high signal-to-noise observations
of the HCN(1--0) and CS(2--1) line profiles toward
selected positions in the dense cores of the clouds
S140, S199, and S235, with the aim of searching for
small-scale structure in molecular cloud cores associated
with regions of formation of high-mass stars.
Some positions were also observed in the C$^{34}$S(2--1)
and H$^{13}$CN(1--0) isotope lines.

In several cases, we found ripples in the line profiles 
of the main isotopes, which can be interpreted
as revealing the presence of a large number of unresolved
small clumps in the telescope beam. We have
used the analytical model [9, 11] to estimate the total
number of small clumps in the telescope beam on
scales $\sim 0.2-0.5$~pc, which can be 
$\sim 2\cdot 10^4-3\cdot 10^5$ 
depending on the source.

Detailed calculations of the excitation of the HCN
and CS lines in a cloud model consisting of a set of
small thermal clumps that are randomly distributed
and move with random velocities have enabled us to
estimate physical properties of clumps toward two
positions in the dense core of S140 and in S199(0,0).
The densities in the clumps are $\sim 3\cdot 10^5-10^6$~cm$^{-3}$,
and their sizes are $\sim (1-3)\cdot 10^{-3}$~pc. The volume
filling factors of the clumps are $\sim 0.03-0.12$. 
To bring the parameters of the HCN and CS profiles into
agreement, we must include an inter-clump gas with
a density no lower than 
$\sim (2-7)\cdot 10^4$~á¬$^{-3}$.

The internal thermal energy of the clump gas
far exceeds the gravitational energy; and the condition
of pressure equilibrium with the inter-clump
gas is likewise not fulfilled. When isolated, such
spherical clumps can undergo long-term oscillations.
The mean lifetime for a set of clumps can be
determined from the mean time between collisions:
$\sim 10^4-10^5$~years. Such structures probably represent
density fluctuatins arising due to turbulence in
regions of formation of high-mass stars.

\begin{acknowledgments}
The authors are grateful to the staff of the Onsala
Observatory for their help with the observations.
Large contribution in data acquisition was done by
untimely gone L.E.B. Johansson. This work was supported
by the Russian Foundation for Basic Research
(project codes 06-02-16317 and 08-02-00628) and
the Basic Research Program of the Division of Physical
Sciences of the Russian Academy of Sciences on
``Extended Objects in the Universe".
\end{acknowledgments}

\newpage

\begin{figure}[t!]
\setcaptionmargin{5mm}
\onelinecaptionsfalse
\includegraphics[width=18cm]{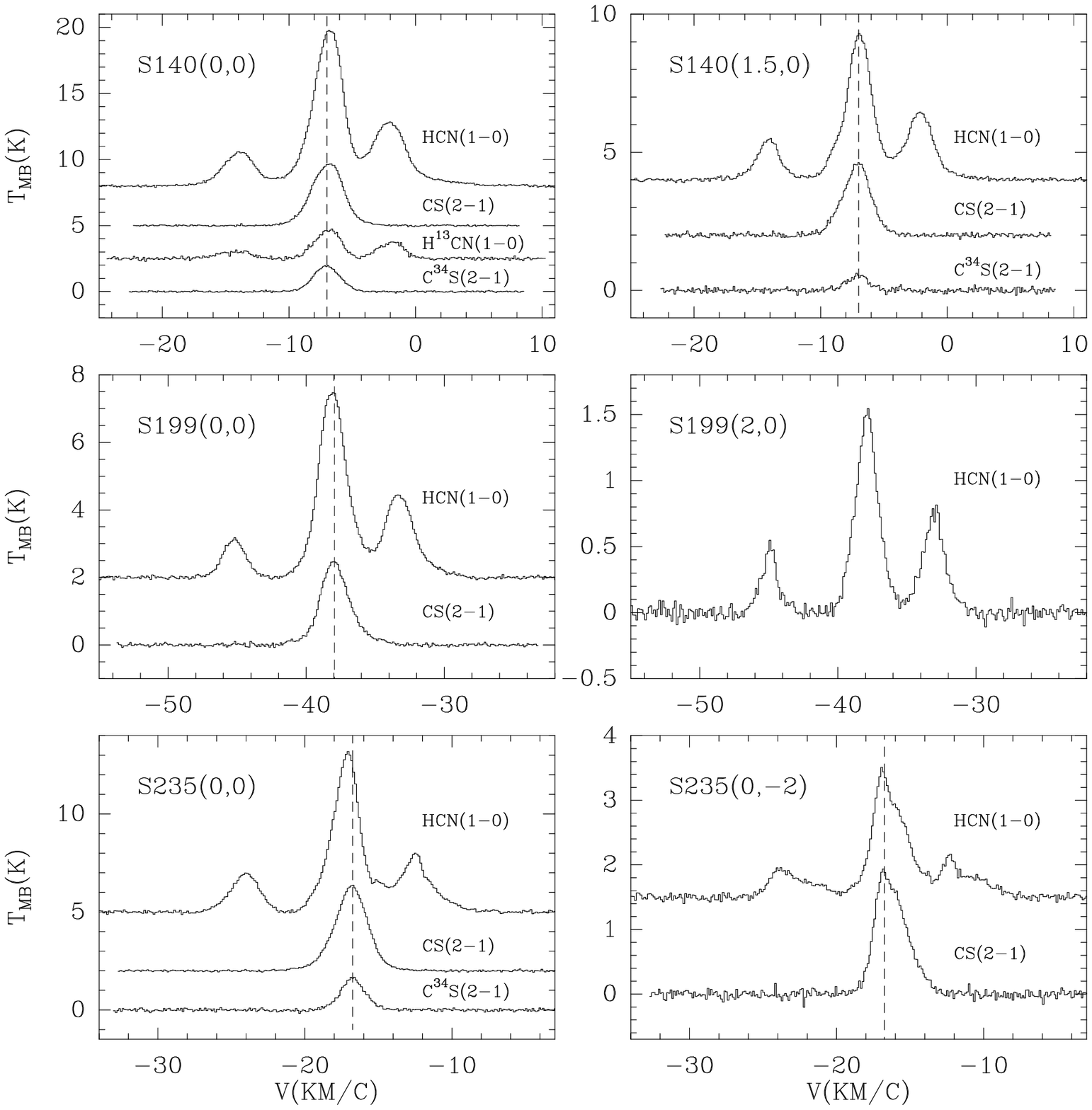}
\captionstyle{normal}
\caption{Measured spectra. The dashed line shows the centers 
of the optically thin C$^{34}$S lines or CS lines.}
\label{fig:spectra}
\end{figure}

\newpage

\begin{figure}[t!]
\setcaptionmargin{5mm}
\onelinecaptionsfalse
\includegraphics[width=15cm]{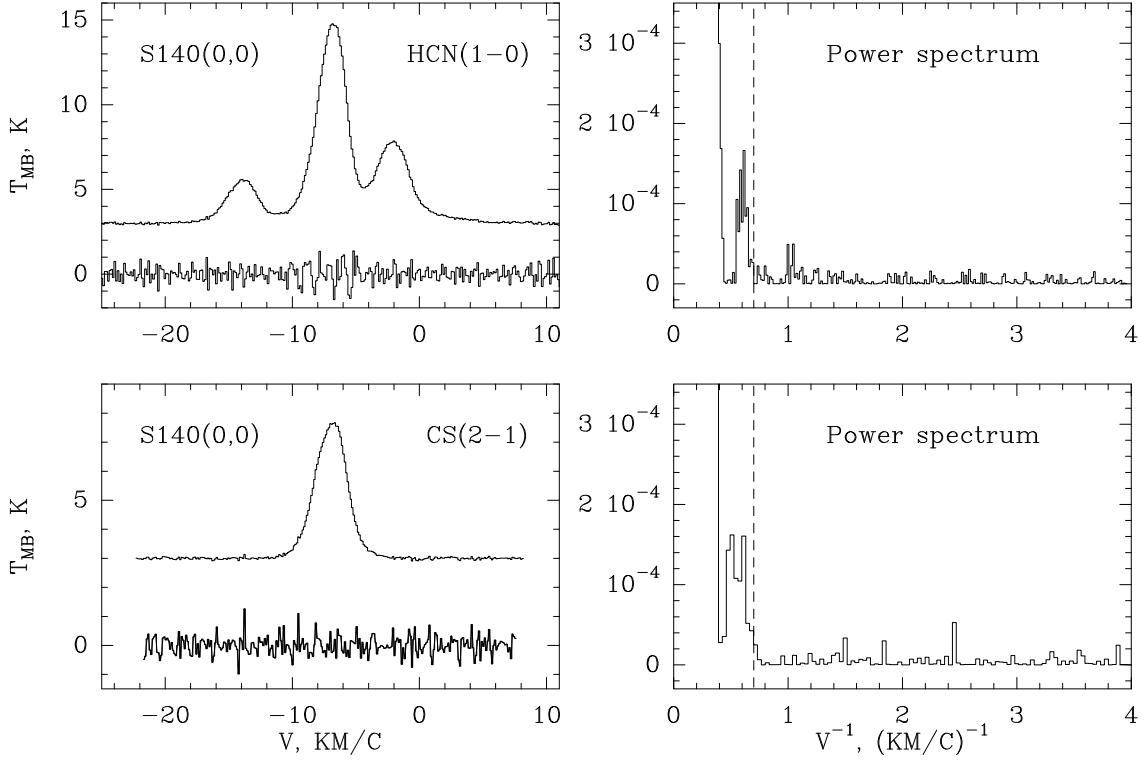}
\captionstyle{flushleft}
\caption{Observed profiles of the HCN(1--0) and CS(2--1) lines together
with the residual fluctuations after rejection of the lowest Fourier harmonics.
The amplitude of the residual fluctuations is magnified by a factor of ten.
The power spectra are shown to the right; the dashed vertical line shows
the cutoff boundary.
}
\label{fig:resid}
\end{figure}

\newpage

\addtocounter{figure}{-1}

\begin{figure}[t!]
\setcaptionmargin{5mm}
\onelinecaptionstrue
\includegraphics[width=15cm]{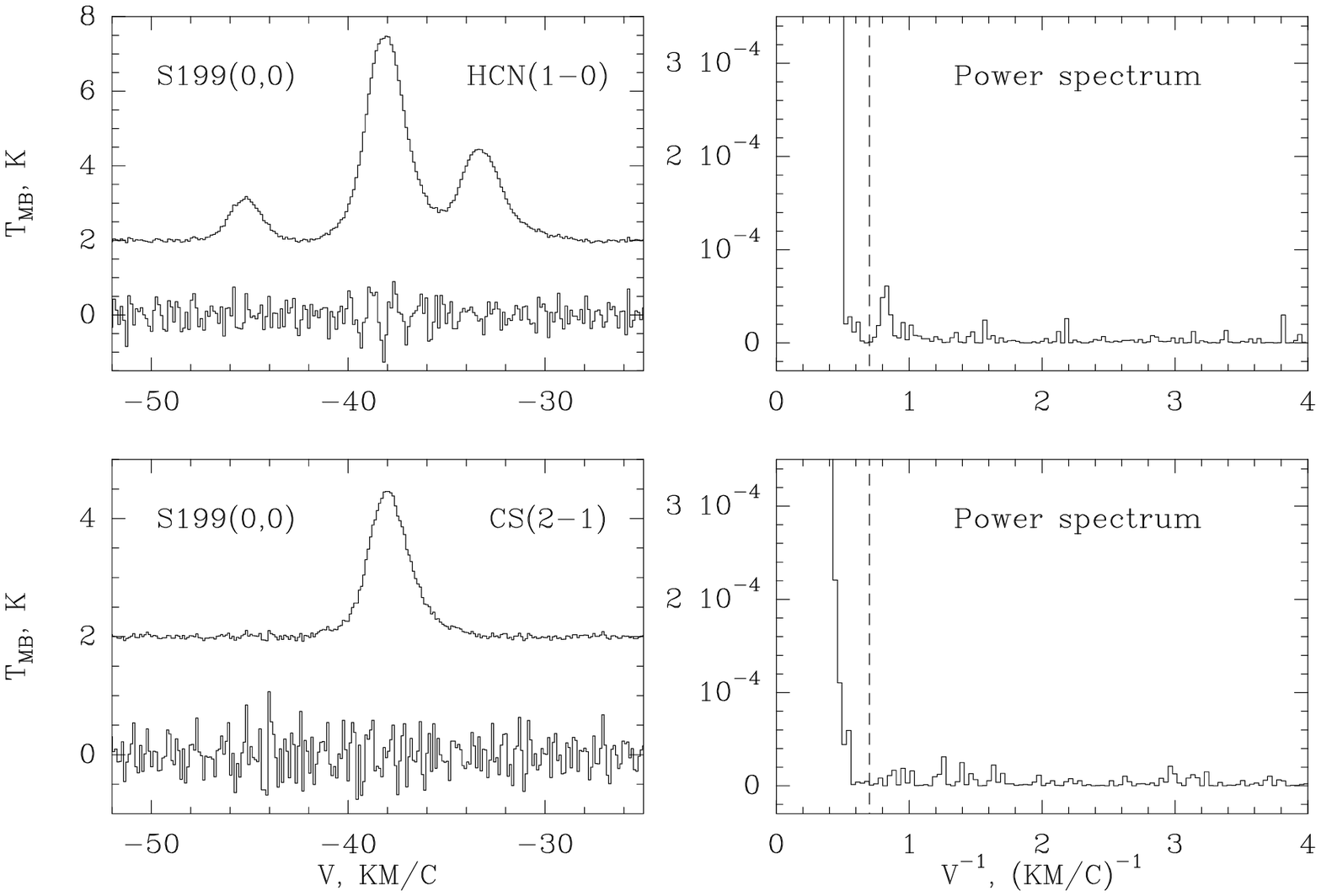}
\captionstyle{normal}
\caption{Contd.}
\end{figure}

\newpage

\addtocounter{figure}{-1}

\begin{figure}[t!]
\setcaptionmargin{5mm}
\onelinecaptionstrue
\includegraphics[width=15cm]{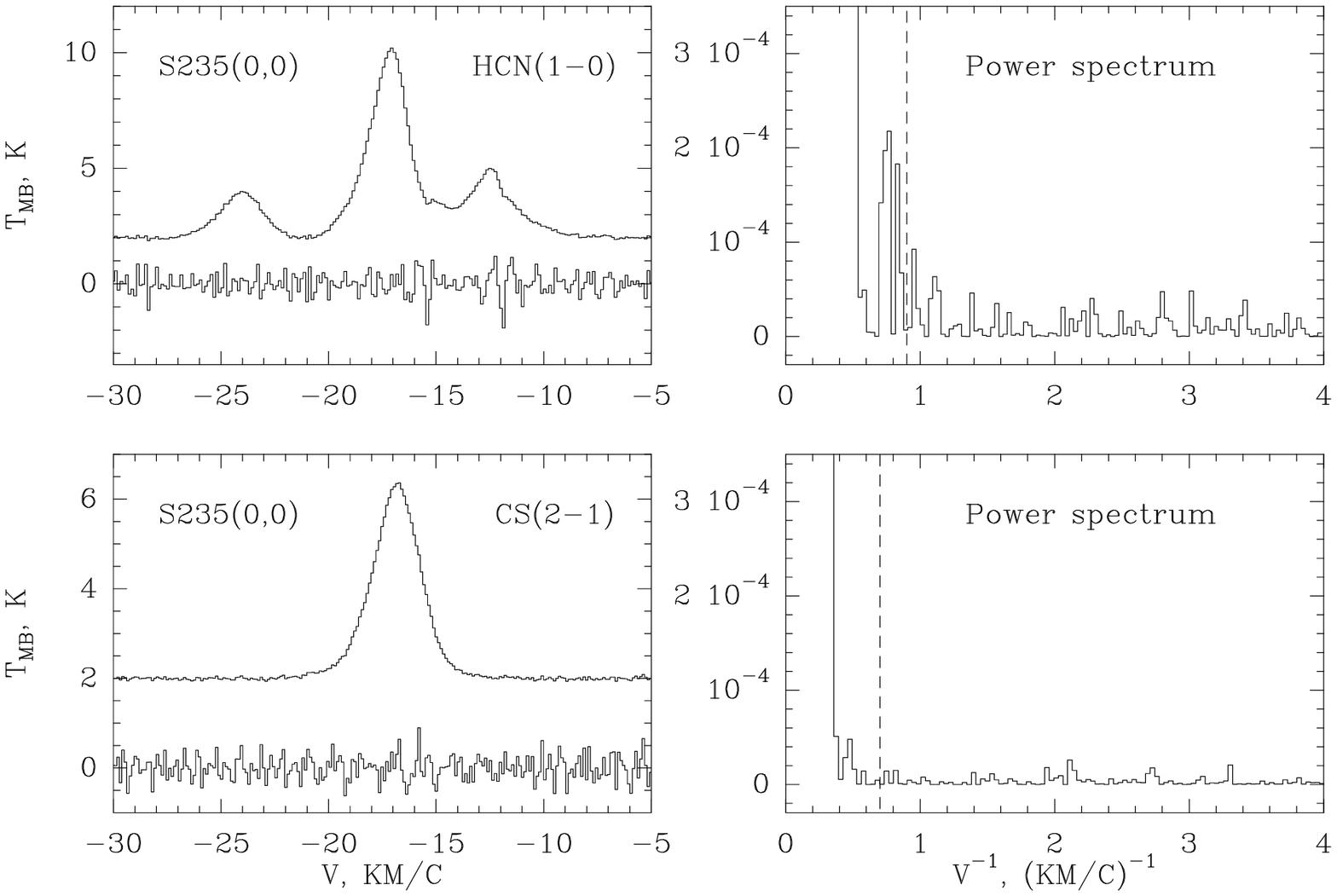}
\captionstyle{normal}
\caption{Contd.}
\end{figure}

\newpage

\begin{figure}[htbp]
\setcaptionmargin{5mm}
\onelinecaptionsfalse

\begin{minipage}[b]{0.48\textwidth}
    \includegraphics[width=\textwidth]{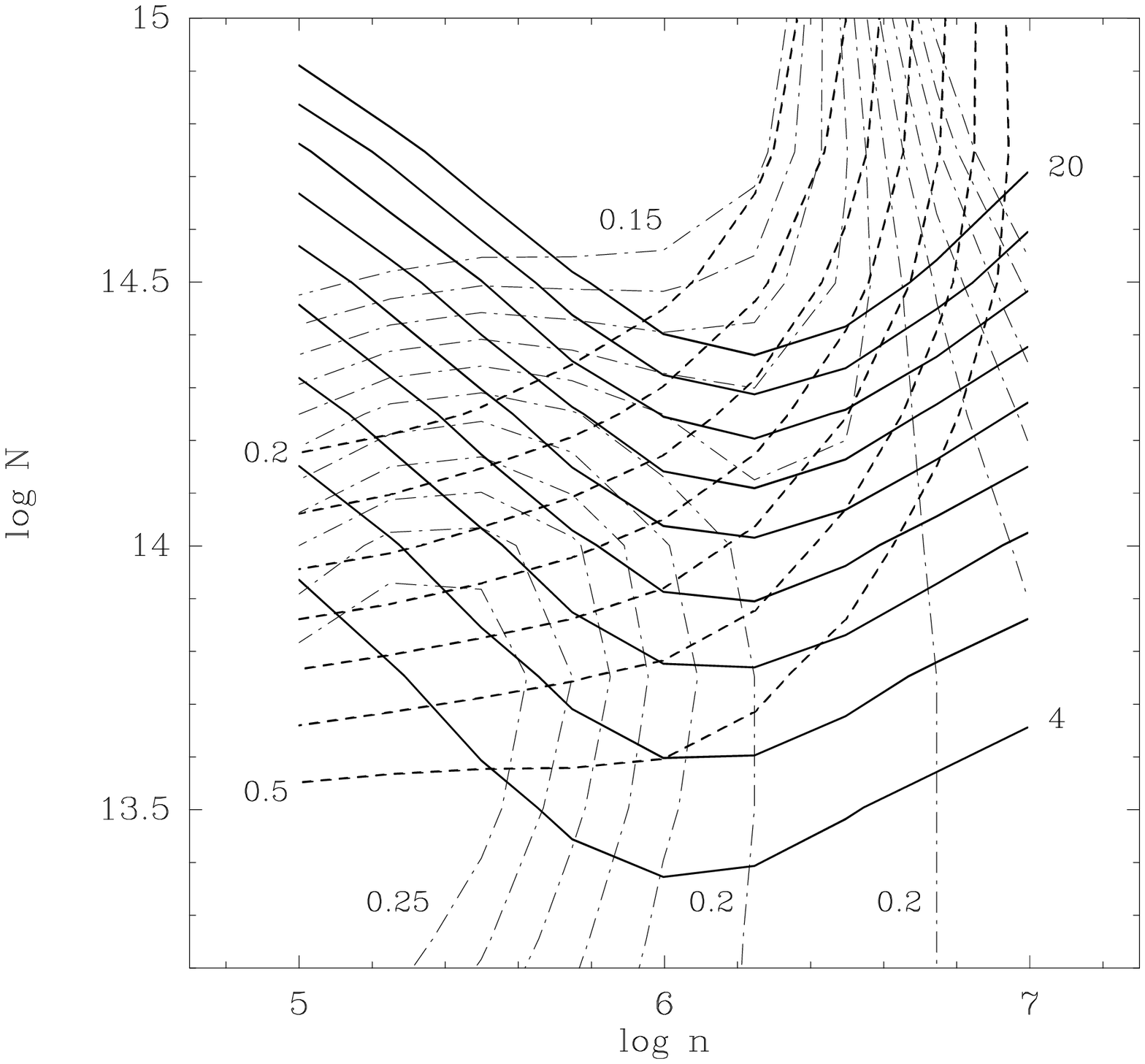}
\end{minipage}
\hfill
\begin{minipage}[b]{0.48\textwidth}
    \includegraphics[width=\textwidth]{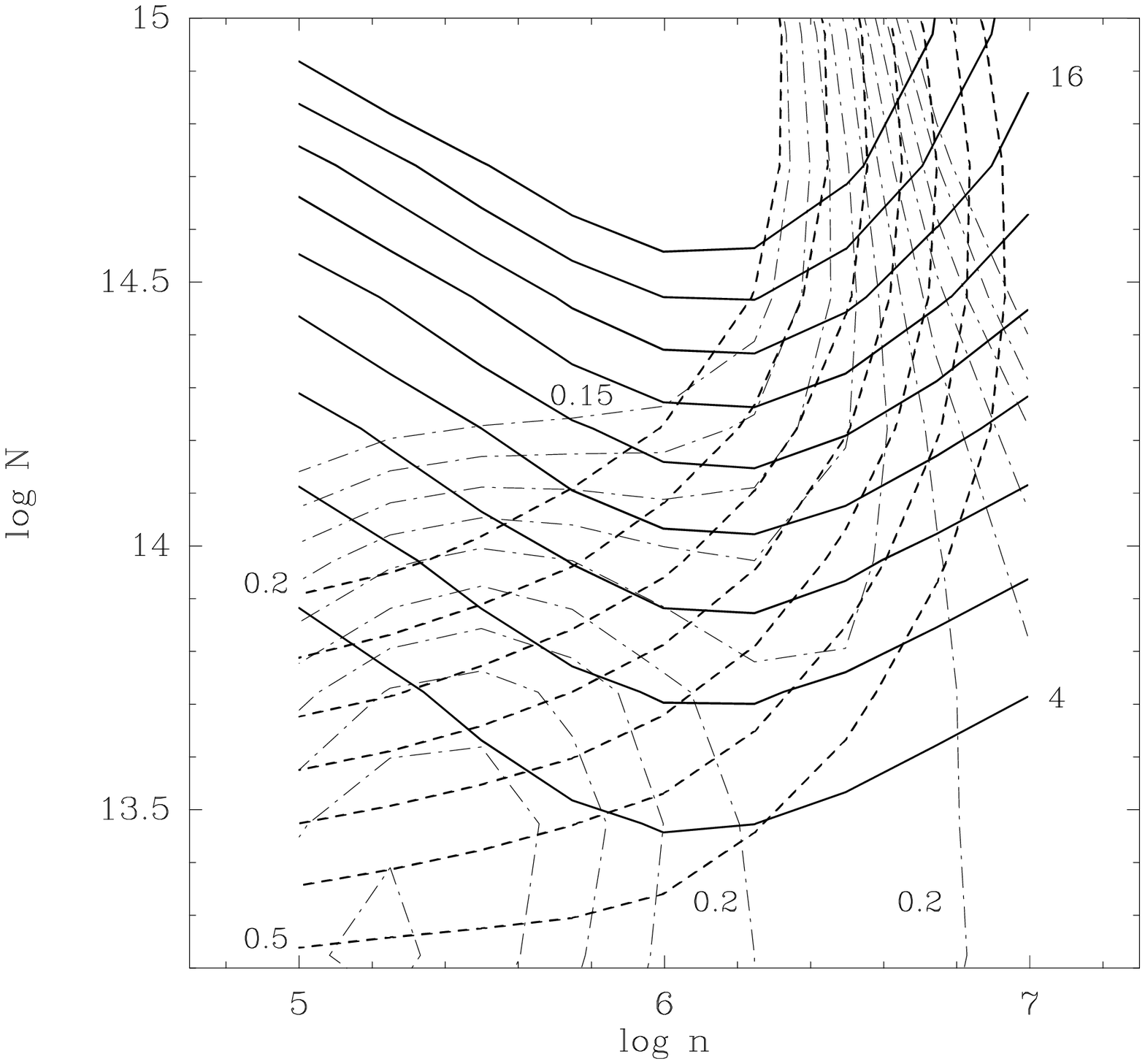}
\end{minipage}

\captionstyle{flushleft}
\caption{Results of calculations of the HCN(1--0) line excitation 
in a model with thermal clumps for two ratios of the volume filling factor
to the clump size.
The contours correspond to the brightness temperatures for the 
$F=2-1$ component (thick solid curves, step 2~K), the relative intensities
of the $F=1-1$ component (dashed curves, step 0.05) and the relative
intensities of the $F=0-1$ component (dash-dot curves, step 0.01).
The range of variation of the values is indicated in the figure.
The physical properties of the model are indicated in the text.
}

\label{fig:iso-model}
\end{figure}

\newpage

\begin{figure}[t!]
\setcaptionmargin{5mm}
\onelinecaptionsfalse
\includegraphics[width=15cm]{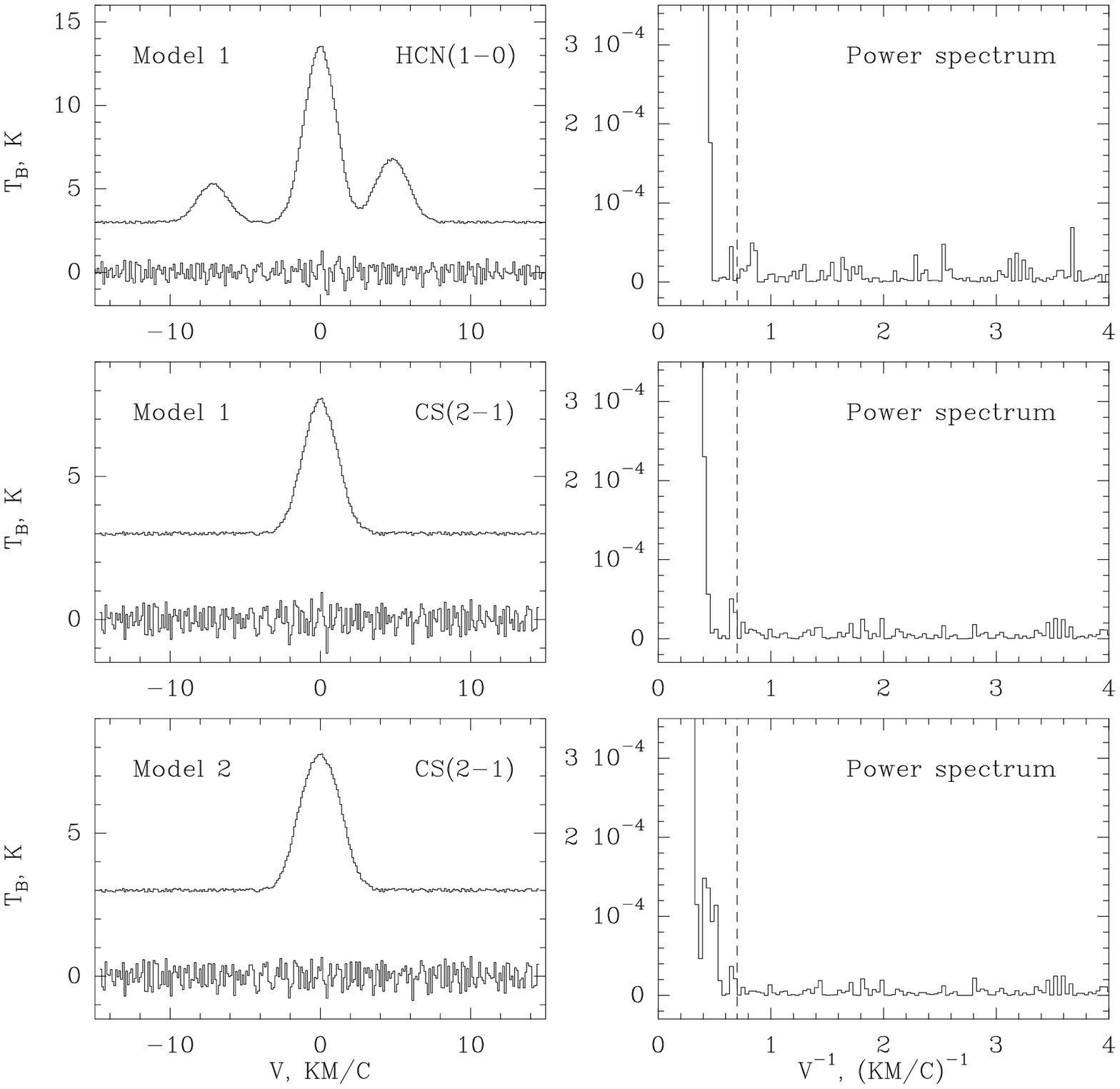}
\captionstyle{flushleft}
\caption{Model profiles of the HCN(1--0) and CS(2--1) lines corresponding 
to those observed in S140(0,0) and obtained for a clumpy model without
(Model 1) and with (Model 2) inter-clump gas. Synthetic noise with dispersion
equal to that of the noise observed outside the line range was added 
to the profiles. The residual noise obtained after filtration of the lowest
Fourier harmonics is shown beneath the model profiles, with its amplitude 
magnified by a factor of ten. The corresponding power spectra for small
amplitudes is shown to the right; the dashed vertical line shows the cutoff
boundary, $0.7$~(km/s)$^{-1}$. The model parameters are given in the text.}
\label{fig:spec-model}
\end{figure}

\newpage

\begin{table}[hbtp]
\setcaptionmargin{0mm}
\onelinecaptionstrue
\captionstyle{flushleft}
\caption{Source list}
\bigskip
\begin{tabular}{|c|c|c|c|c|l|c|}\hline\noalign{\smallskip}
Source          & $\alpha$(1950) & $\delta$(1950) & D      & Resolution\\
                  &${\rm (^h)\  (^m)\  (^s)\ }$
                                   &$(^o)$  $(^{\prime})$  $(^{\prime\prime}$)
                                                    & (kpc)  & (pc)\\
\noalign{\smallskip}\hline\noalign{\smallskip}
S140    &  22 17 41.3 &  63 03 40  & 0.9 \cite{FB84}     & 0.17--0.19 \\
S199    &  02 57 35.6 &  60 17 22  & 2.2 \cite{Snell88}  & 0.41--0.46 \\
S235    &  05 37 31.8 &  35 40 18  & 2.3 \cite{BB93}     & 0.43--0.48 \\
\noalign{\smallskip}\hline\noalign{\smallskip}
\end{tabular}

\label{table:list}
\end{table}

\newpage

\begin{table}[hbtp]
\small
\setcaptionmargin{0mm}
\onelinecaptionstrue
\captionstyle{flushleft}
\caption{Parameters of the observed lines}
\bigskip
\begin{tabular}{|c|c|c|c|c|c|}\hline\noalign{\smallskip}
Source        & Line & $T_{\rm R}$(K) & $R_{\rm 12}$, $R_{\rm 02}$ & $\Delta V$(km/s) & $\tau_{\rm eff}$\\
\noalign{\smallskip}\hline\noalign{\smallskip}
S140 (0',0')    & HCN(1--0)        &10.9(0.1)$^1$ &0.38, 0.20    & 2.55(0.01) & 0.42(0.03) \\
                & H$^{13}$CN(1--0) & 1.09(0.02)   &0.57, 0.24    & 2.37(0.04) &            \\
                & CS(2--1)         & 4.70(0.02)   &              & 2.90(0.01) & 1.05(0.05) \\
                & C$^{34}$S(2--1)  & 0.97(0.01)   &              & 2.44(0.02) &            \\
S140 (1.5',0')  & HCN(1--0)        & 5.06(0.03)   &0.47, 0.28    & 2.50(0.02) & 0.9(0.2)   \\
                &                  &              &              & 2.14(0.05)$^2$ \\
                & CS(2--1)         & 2.52(0.02)   &              & 2.61(0.02) & 1.6(0.4)   \\
                & C$^{34}$S(2--1)  & 0.27(0.01)   &              & 2.0(0.1)   &            \\
\noalign{\smallskip}\hline\noalign{\smallskip}
S199    (0',0') & HCN(1--0)        & 5.44(0.03)   &0.43, 0.20    & 2.40(0.01) & 1.6(0.3)   \\
                &                  &              &              & 1.87(0.06)$^2$ \\
                & CS(2--1)         & 2.4(0.02)    &              & 2.45(0.02) &            \\
S199    (2',0') & HCN(1--0)        & 1.45(0.02)   &0.50, 0.31    & 2.00(0.03) &            \\
\noalign{\smallskip}\hline\noalign{\smallskip}
S235  (0',0')   & HCN(1--0)        & $\sim 8^3$   &              & 2.29(0.03)$^2$  &       \\
                & CS(2--1)         & 4.21(0.02)   &              & 2.59(0.01)& 1.9(0.1)    \\
                & C$^{34}$S(2--1)  & 0.79(0.01)   &              & 1.95(0.03)&             \\
S235  (0',--2') & HCN(1--0)        & $\sim 2^3$   &              &           &             \\
                & CS(2--1)         & $\sim 1.8^3$ &              & $\sim 2.5^3$   &        \\
\noalign{\smallskip}\hline\noalign{\smallskip}
\end{tabular}

\begin{flushleft}
$^1$ -- After subtracting the high-velocity component.\\
$^2$ -- Width of the $F=0-1$ component.\\
$^3$ -- More than one component in the profiles.
\end{flushleft}

\label{table:param}
\end{table}

\newpage

\begin{table}[hbtp]
\small
\setcaptionmargin{0mm}
\onelinecaptionsfalse
\captionstyle{flushleft}
\caption{Residual fluctuations in the line profiles and 
the total number of clumps in the telescope beam calculated  
in the analytical model}
\begin{tabular}{|c|c|c|c|c|c|}\hline\noalign{\smallskip}
Source        & Line & $\Delta T_{\rm N}$(K)& $\Delta T_{\rm R}(K)$ & $T_{\rm KIN}$(K) & $N_{\rm tot}$ \\
\noalign{\smallskip}\hline\noalign{\smallskip}
S140 (0',0')    & HCN(1--0)        & 0.042 & 0.055      & 30 \cite{Mal05}   & $\sim 2.6\cdot 10^5$ \\
                & H$^{13}$CN(1--0) & 0.037 &            &                   & \\
                & CS(2--1)         & 0.035 & $<0.015$   &                   & $>4.2\cdot 10^5$ \\
                & C$^{34}$S(2--1)  & 0.018 &            &                   & \\
S140 (1.5',0')  & HCN(1--0)        & 0.035 & 0.055      & 39 \cite{Mal05}   & $\sim 2.7\cdot 10^4$ \\
                & CS(2--1)         & 0.051 & $<0.02$    &                   & $>2.5\cdot 10^4$  \\
                & C$^{34}$S(2--1)  & 0.035 &            &                   & \\
\noalign{\smallskip}\hline\noalign{\smallskip}
S199    (0',0') & HCN(1--0)        & 0.032 & 0.053      & 28 \cite{Zin97}   & $\sim 1.7\cdot 10^4$ \\
                & CS(2--1)         & 0.031 & 0.016      &                   & $\sim 3.9\cdot 10^4$ \\
\noalign{\smallskip}\hline\noalign{\smallskip}
S235  (0',0')   & HCN(1--0)        & 0.050 & 0.056      & 40 \cite{Zin98}   & $\sim 2.3\cdot 10^4$ \\
                & CS(2--1)         & 0.027 & 0.025      &                   & $\sim 4.0\cdot 10^4$ \\
                & C$^{34}$S(2--1)  & 0.026 &            &                   & \\
      \noalign{\smallskip}\hline\noalign{\smallskip}
\end{tabular}

\label{table:ntot}
\end{table}

\newpage

\begin{table}[hbtp]
\small
\setcaptionmargin{0mm}
\onelinecaptionstrue
\captionstyle{flushleft}
\caption{Physical parameters of clumps obtained from model calculations}
\begin{tabular}{|c|c|c|c|c|c|c|c|c|}\hline\noalign{\smallskip}
Source & $N_{\rm tot}$ & $f$ & $d$   & $n$         & $n_{\rm ic}$ & $M$           & $M_{\rm vir}$ & $P_{\rm ext}/P_{\rm int}$ \\
         &               &     & (pc)  & (cm$^{-3}$) & (cm$^{-3}$)  & ($M_{\odot}$) & ($M_{\odot})$ & \\
\noalign{\smallskip}\hline\noalign{\smallskip}
S140 (0',0')   & $\sim 2.8\cdot 10^5$   & $4.2\cdot 10^{-2}$ & $\sim 10^{-3}$        & $8.5\cdot 10^5$ & $5\cdot 10^4$ & $\sim 3\cdot 10^{-5}$ & 0.062 & 0.2\\
S140 (1.5',0') & $\sim 3.6\cdot 10^4$   & $1.2\cdot 10^{-1}$ & $\sim 3\cdot 10^{-3}$ & $3.4\cdot 10^5$ & $7\cdot 10^4$ & $\sim 3\cdot 10^{-4}$ & 0.25  & 0.5\\
S199 (0',0')   & $\sim 1.1\cdot 10^5$   & $2.5\cdot 10^{-2}$ & $\sim 3\cdot 10^{-3}$ & $1.2\cdot 10^6$ & $2\cdot 10^4$ & $\sim 10^{-3}$        & 0.19  & 0.06\\
\noalign{\smallskip}\hline\noalign{\smallskip}
\end{tabular}

\label{table:physpar}
\end{table}

\end{document}